\documentclass[a4paper,twocolumn,11pt]{quantumarticle}

\pdfoutput=1

\usepackage[utf8]{inputenc}
\usepackage[T1]{fontenc}
\usepackage{graphicx}
\usepackage{amsmath, amssymb}
\usepackage{physics}
\usepackage[numbers,sort&compress]{natbib}
\usepackage{algorithm}
\usepackage{algpseudocode}
\usepackage{hyperref}
\usepackage{caption}
\usepackage{subcaption}
\usepackage{needspace}
\usepackage{xurl}

\graphicspath{{results/}}
\DeclareGraphicsExtensions{.png,.pdf,.jpg}

\title{Geometric-Phase (Pancharatnam--Berry) Correction for Time-Bin Photonic Qudits: A Calibration and Feed-Forward Algorithm}

\author{Ryan Rae-Cheng Wee}
\affiliation{Monash Science School, Melbourne, Australia}

\author{Josef Bruzzese}
\orcid{0009-0004-9993-5953}

\date{August 2025}

\begin{document}

\maketitle

\begin{abstract}
We develop a geometric-phase framework for time-bin photonic qudits and propose a practical calibration and feed-forward algorithm that separates and compensates geometric (Pancharatnam--Berry), dynamical, and technical phase contributions. Working directly in the time-bin basis, we impose a parallel-transport gauge so that geometric phases appear as experimentally identifiable interferometric offsets, while all phase contributions enter a bin-resolved diagonal transformation. We model state preparation by cascaded unbalanced Mach--Zehnder interferometers and give closed-form amplitudes for arbitrary splitting ratios and phases, clarifying that single-port monitoring requires post-selection and renormalization. We provide an interferometric tomography recipe (adjacent-bin scans with a Fourier-basis cross-check) and a multi-mode numerical case study that separates total, dynamical, and geometric phases and demonstrates accurate feed-forward compensation. The protocol uses only standard components (tunable UMZIs, phase shifters/EOMs, and single-photon detectors) and routine phase sweeps, making it immediately implementable for small to moderate dimensions ($d\lesssim 10$) and providing a scalable pathway to phase-stable high-dimensional temporal encoding for quantum communication and photonic processing.
\end{abstract}

\section{Introduction}

Understanding and utilizing geometric phases (Berry phases and their generalizations) is of wide significance in quantum mechanics, ranging from condensed matter systems to quantum computation and communication~\cite{Xiao2010,Ghahari2017,Nagaosa2009}. In photonic platforms, geometric phases can be engineered through polarization optics (Pancharatnam--Berry phase) or through controlled parameter loops in interferometric networks, and have been proposed as resources for robust control and fault-tolerant operations.

Time-bin encoding is a leading approach for photonic quantum communication because it is naturally compatible with fiber links, is resilient to polarization drift, and supports high-dimensional encodings (qudits) for increased channel capacity and improved noise tolerance~\cite{Tittel2000,Marcikic2003,Brendel1999,Bouchard2022}. Recent proposals have also highlighted the use of time-bin photonic qudits as compact high-dimensional processors, including schemes in which an \(N\)-qubit logical register is encoded in a single \(2^N\)-dimensional time-bin qudit using \(O(N)\) linear-optical elements~\cite{Delteil2024}. These architectures make scalable, bin-resolved phase calibration increasingly important, since phase errors across many temporal modes directly affect preparation, manipulation, and measurement. In practice, time-bin qudits are typically generated and analysed using unbalanced Mach--Zehnder interferometers (UMZIs), cascaded interferometric trees, or related delay-line architectures, all of which naturally introduce phase accumulation across multiple optical paths.

The central problem addressed here is that, as the dimension $d$ increases, experimentally observed relative phases between time bins unavoidably mix (i) \emph{dynamical} phases from propagation and dispersion and any effective Hamiltonian evolution, (ii) \emph{geometric} (Pancharatnam--Berry) phases associated with controlled parameter loops, and (iii) \emph{technical} phases from slow drifts and control offsets. In time-bin qudits these contributions are accessible only through interference fringes, so a systematic, bin-resolved procedure to separate (or at least consistently estimate) and compensate them has not been addressed in a unified way for high-dimensional temporal encodings.

We present a systematic solution: a parallel-transport gauge in the time-bin basis that isolates the geometric contribution as an interferometric offset, together with a calibration routine that infers bin-to-bin relative phases and a diagonal feed-forward correction that compensates the full phase budget. This makes geometric phase not only measurable but also correctable with standard laboratory components, enabling phase-stable high-dimensional quantum communication and photonic processing.

\subsection{Contributions}\label{subsec:contributions}

Our contributions are as follows.
\begin{enumerate}
    \item \emph{Geometric-phase framework in the time-bin basis.} We formulate the Pancharatnam--Berry phase directly on time-bin amplitudes and adopt a parallel-transport gauge in which geometric contributions appear as stable interferometric offsets, distinguishable in principle from the dynamical and technical contributions that share the same diagonal structure (Sec.~\ref{sec:prelim} and Sec.~\ref{sec:tb}).

    \item \emph{State-preparation model with closed-form amplitudes.} We model generation by cascaded unbalanced Mach--Zehnder interferometers (UMZIs) and give closed-form amplitudes $\alpha_j$ for arbitrary splitting ratios $\eta_k$ and arm phases $\phi_k$, clarifying that single-port monitoring corresponds to a post-selected, renormalised conditional state rather than the unconditional output (Sec.~\ref{subsec:tb-generation}--\ref{subsec:tb-postselection}).

    \item \emph{Calibration and tomography protocol.} We give an interferometric recipe that recovers the full set of adjacent-bin relative phases $\{\Delta\theta_j\}$ and their visibilities $V_{j,j+1}$ from sinusoidal fringe scans, with an approximate Fourier-basis measurement as an over-constraining cross-check on the reconstructed state and a coherence diagnostic (Sec.~\ref{sec:calibration}); a balanced two-stage qutrit illustrates the procedure analytically (Sec.~\ref{subsec:qutrit-example}).

    \item \emph{Diagonal feed-forward correction.} From the calibrated $\{\Delta\theta_j\}$ we construct a diagonal feed-forward unitary $D_{\mathrm{corr}}$ that cancels the full phase budget---geometric, dynamical, and technical---up to a global phase, implementable with programmable per-bin phase shifters (EOMs or pulse shapers) and routine phase sweeps (Sec.~\ref{sec:correction}).

    \item \emph{Numerical case study.} On a four-bin qudit formed by two decoupled spin-$\tfrac12$ systems we separate the total, dynamical, and geometric phases within numerical precision and verify that the inferred per-bin phases define a correction restoring the target bin phases up to a global phase (Sec.~\ref{sec:correction}).
\end{enumerate}

\paragraph{Main assumptions.} The protocol targets small to moderate dimensions ($d \lesssim 10$) and adiabatic or quasi-adiabatic evolutions. We assume (a) near-orthogonality of time bins, $\Delta t \gg \sigma_{\mathrm{pulse}} + \sigma_{\mathrm{jitter}}$; (b) access to tunable UMZIs and per-bin phase shifters (EOMs or pulse shapers); (c) effectively lossless interferometers and standard single-photon detection; and (d) negligible higher-order effects---e.g.\ uncompensated dispersion---over the calibration window. The numerical case study additionally assumes lossless, dispersion-free propagation; the regime of validity and these idealisations are discussed in Sec.~\ref{sec:tb}.

\paragraph{Code availability.} The numerical routines used to generate the figures and to validate the
calibration/correction procedure are publicly available. Versioned code
(tag \texttt{v1.0.1}, commit \texttt{b4baaf1}) is hosted on
\href{https://github.com/97RyanW/Geometric-Phase-Correction}{GitHub}
and archived on Zenodo with
DOI~\href{https://doi.org/10.5281/zenodo.20036053}{10.5281/zenodo.20036053}.

\section{Preliminaries}\label{sec:prelim}

\subsection{Geometric phase and parallel transport}
A physical pure state is a ray, so an evolving state $\ket{\psi(t)}$ is defined only up to a gauge choice $\ket{\psi(t)}\mapsto e^{i\chi(t)}\ket{\psi(t)}$. For a general evolution from $0$ to $T$, the Pancharatnam total phase is $\beta(T)=\arg\!\braket{\psi(0)}{\psi(T)}$, while the dynamical phase is $\phi_{\mathrm{dyn}}(T)=-(1/\hbar)\int_0^T \mel{\psi(t)}{H(t)}{\psi(t)}\,dt$. The geometric (Pancharatnam--Berry) phase is the gauge-invariant remainder~\cite{Pancharatnam1956,AharonovAnandan1987,Cohen2019}
\begin{equation}
\begin{aligned}
  \gamma(T) &= \beta(T)-\phi_{\mathrm{dyn}}(T) \\
  &= \arg\!\braket{\psi(0)}{\psi(T)} \\
  &\quad +\frac{1}{\hbar}\int_0^T \mel{\psi(t)}{H(t)}{\psi(t)}\,dt .
\end{aligned}
  \label{eq:geom-general}
\end{equation}
For an instantaneous eigenstate that undergoes a cyclic adiabatic parameter loop $\boldsymbol{\lambda}(t)$, this reduces to the Berry phase~\cite{Berry1984}
\begin{equation}
  \gamma_{\mathrm{B}}
  = i\oint_{\mathcal{C}}
  \mel{n(\boldsymbol{\lambda})}{\nabla_{\boldsymbol{\lambda}}}{n(\boldsymbol{\lambda})}\cdot d\boldsymbol{\lambda}.
  \label{eq:berry-prelim}
\end{equation}
In later sections we adopt a parallel-transport convention in the time-bin basis so that geometric contributions become experimentally identifiable offsets in standard interferometric phase sweeps.

\subsection{Time-bin basis}
Time-bin encoding uses well-separated temporal modes as a computational basis for a single photon~\cite{Brendel1999}. We write $\{\ket{t_j}\}_{j=0}^{d-1}$ for time bins separated by $\Delta t$ and assume near-orthogonality when $\Delta t$ is large compared to the pulse duration and detector jitter. A general pure time-bin qudit is
\begin{equation}
  \ket{\psi}=\sum_{j=0}^{d-1}\alpha_j\ket{t_j},
  \qquad
  \sum_{j=0}^{d-1}|\alpha_j|^2=1.
  \label{eq:tb-prelim}
\end{equation}
Only relative phases between amplitudes $\alpha_j$ affect interference, motivating the bin-resolved phase estimation and correction developed below.

\section{Time bin qudits}\label{sec:tb}
Time-bin photonic qudits encode information in discrete temporal modes of a single photon and are typically prepared and analysed with unbalanced Mach--Zehnder interferometers (UMZIs) and calibrated delays in fiber or integrated photonics~\cite{Yu2025}. As the dimension $d$ increases, phase control becomes the dominant bottleneck: each time bin accumulates phase contributions from many interferometric paths, and these phases directly determine the interference fringes used for state analysis and downstream processing.

\paragraph{Problem statement.}
The experimentally observed relative phases between time bins generally mix three contributions: (i) \emph{dynamical} phases from propagation (path length, dispersion) and any effective Hamiltonian evolution; (ii) \emph{geometric} (Pancharatnam--Berry) phases associated with controlled parameter loops; and (iii) \emph{technical} phases from slow drifts and control electronics. In time-bin qudits, these contributions add and are accessed only through interference. A systematic, bin-resolved procedure to separate (or at least consistently estimate) these contributions and compensate them via feed-forward is essential for high-dimensional temporal encodings and is the focus of this work.

\paragraph{Why it matters.}
On the theory side, operational isolation of geometric phase elevates it from a conceptual quantity to a resource for holonomic control. On the practical side, phase-stable bin-resolved interference is necessary for error-resilient high-dimensional quantum communication and scalable photonic processing, where small drifts otherwise compound across many time bins.

\paragraph{Core idea of the solution.}
We work in a parallel-transport gauge in the time-bin basis so that geometric phases appear as identifiable offsets in standard interferometric phase sweeps. We then infer a consistent set of relative phases from adjacent-bin fringes and apply a diagonal feed-forward correction (programmable per-bin phase shifts) that compensates geometric, dynamical, and technical phases together, restoring the intended target state up to a global phase.

\paragraph{Applicability and assumptions.}
We target small to moderate dimensions ($d\lesssim 10$) and adiabatic or quasi-adiabatic evolutions, assuming: (i) near-orthogonality of time bins ($\Delta t \gg$ pulse width $+$ detector jitter), (ii) access to tunable UMZIs and per-bin phase shifters (EOMs or pulse shapers), and (iii) negligible higher-order effects over the calibration window (e.g.\ uncompensated dispersion). We also assume effectively lossless interferometers and standard single-photon detection.

\subsection{Time-bin state model and orthogonality}\label{subsec:tb-model}
We represent the time-bin basis as $\{\ket{t_j}\}_{j=0}^{d-1}$ with separation $\Delta t$. Under the near-orthogonality assumption,
\begin{equation}
\braket{t_j}{t_k}\approx \delta_{jk},\qquad
\Delta t \gg \sigma_{\rm pulse}+\sigma_{\rm jitter},
\label{eq:orthogonality}
\end{equation}
where $\sigma_{\rm pulse}$ is the pulse temporal width and $\sigma_{\rm jitter}$ the detector timing jitter. A general pure time-bin qudit is
\begin{equation}
\ket{\psi}=\sum_{j=0}^{d-1}\alpha_j\ket{t_j},\qquad \sum_{j=0}^{d-1}|\alpha_j|^2=1.
\label{eq:tb-state}
\end{equation}
Throughout, global phase is physically irrelevant, and only relative phases between $\alpha_j$ affect interference.

\subsection{Generation via cascaded UMZIs}\label{subsec:tb-generation}
A common generation method uses a cascade of $m$ UMZIs, each splitting an incoming temporal mode into an early/late superposition. For stage $k$, let $\eta_k$ be the power splitting ratio and $\phi_k$ the relative phase between arms. The single-photon transformation can be written as
\begin{equation}
\ket{t}\mapsto \sqrt{\eta_k}\,\ket{t}+\sqrt{1-\eta_k}\,e^{i\phi_k}\ket{t+\Delta t}.
\label{eq:umzi-stage}
\end{equation}
After $m$ stages, up to $d=m+1$ distinct time bins appear. For arbitrary $\{\eta_k,\phi_k\}$ one obtains closed-form amplitudes $\alpha_j$ by summing path contributions; in the balanced case $\eta_k=\tfrac12$ the tree yields uniform amplitudes with phases determined by the $\phi_k$.

\subsection{Post-selection and renormalization}\label{subsec:tb-postselection}
Single-port monitoring of a multi-stage interferometric tree typically requires post-selection, since not all paths exit the chosen port. Operationally, one measures a conditional state on the monitored port and renormalizes the amplitudes. This is important for interpreting tomography and calibration: measured probabilities and fringes correspond to the post-selected state, not the unconditional output distribution.

\subsection{Fourier transform basis and interferometric analysis}\label{subsec:tb-fourier}
For state analysis and cross-checks, it is useful to consider the discrete Fourier transform (DFT) basis. Define
\begin{equation}
F_d\ket{t_j}=\frac{1}{\sqrt{d}}\sum_{\ell=0}^{d-1}e^{2\pi i j\ell/d}\ket{t_\ell},
\label{eq:dft}
\end{equation}
which maps the time-bin basis to a mutually unbiased basis. In practice, approximate implementations use multi-delay interferometers and phase shifters. Measuring in both the computational basis (time-resolved detection) and one or more interferometric bases provides an over-constrained reconstruction of the state and is valuable for validating phase calibration.

\subsection{Geometric phase in the time-bin basis}\label{subsec:tb-geomphase}
To connect geometric phase to time-bin interferometry, we impose a parallel-transport gauge on bin amplitudes. For a (not necessarily eigenstate) evolution $\ket{\psi(t)}$, the geometric phase over $[0,T]$ is given by Eq.~\eqref{eq:geom-general}. In the adiabatic cyclic case, it reduces to the Berry phase in Eq.~\eqref{eq:berry-prelim}. In time-bin circuits, the relevant observables are bin-to-bin relative phases inferred from fringes; our approach uses gauge fixing and adjacent-bin interference to extract these phases in a way that isolates geometric contributions as stable interferometric offsets, while keeping the correction procedure agnostic to the microscopic origin of the phases (geometric vs dynamical vs technical).

\section{Calibration and phase separation}\label{sec:calibration}
This section makes the phase-separation problem operational. Because time-bin experiments access phases only through interference, we frame calibration as the task of recovering a consistent set of \emph{relative} phases between bins and then constructing a feed-forward correction that cancels them. At each step we clarify the goal, the measured observable, and the inferred quantity.

\subsection{Phase error model and calibration procedure}\label{subsec:tb-calibration}
In practice, each time bin acquires a total phase
\begin{equation}
  \theta_j = \phi^{\mathrm{dyn}}_j + \gamma_j + \phi^{\mathrm{tech}}_j,
  \quad j=0,\dots,d-1,
  \label{eq:phase-budget}
\end{equation}
where $\phi^{\mathrm{dyn}}_j$ is dynamical (e.g., path length, dispersion, effective Hamiltonian evolution), $\gamma_j$ is geometric (Berry for adiabatic loops, Pancharatnam--Berry in general), and $\phi^{\mathrm{tech}}_j$ accounts for technical offsets and drifts (EOM drive, temperature, timing skew). The effective transformation on the state \eqref{eq:tb-state} is therefore a diagonal unitary
\begin{equation}
  U_{\mathrm{tot}} = \mathrm{diag}\!\left(e^{i\theta_0},\dots,e^{i\theta_{d-1}}\right).
  \label{eq:Utot}
\end{equation}

\noindent\textbf{Goal.} Recover experimentally relevant adjacent-bin relative phases
\[
\Delta\theta_j := \theta_{j+1}-\theta_j,\qquad j=0,\dots,d-2,
\]
which together determine all per-bin phases up to a global reference. Each $\Delta\theta_j$ includes geometric, dynamical, and technical contributions. Once $\{\Delta\theta_j\}$ is known, a diagonal feed-forward correction cancels the full phase budget at once; if an independent dynamical model is available, one may additionally subtract it to quantify geometric contributions.

\medskip
\noindent\textbf{Step 1: adjacent-bin interferometric scans.}
For each adjacent pair $(j,j{+}1)$, analyse with a UMZI of delay $\Delta t$ and scan a tunable analysis phase $\varphi$. The single-port detection probability is
\begin{equation}
\begin{aligned}
  P_{j,j+1}(\varphi) &= \tfrac{1}{2}\!\left(p_j + p_{j+1}\right) \\
  &\quad + \Re\!\left[e^{i\varphi}\rho_{j,j+1}\right],\quad p_j=\rho_{jj},
\end{aligned}
  \label{eq:adj-fringe}
\end{equation}
where $\rho_{j,j+1}=\alpha_j\alpha_{j+1}^*$ for a pure state. Fitting the fringe yields (i) a \emph{phase offset} $\arg\rho_{j,j+1}$, which equals the measured relative phase $\Delta\theta_j$ up to a fixed reference choice, and (ii) a \emph{visibility} $V_{j,j+1}=2|\rho_{j,j+1}|/(p_j+p_{j+1})$, which quantifies coherence. High visibility indicates that the relative phase is well defined and that the phase estimate is reliable; reduced visibility flags mode mismatch, loss of coherence, or imperfect delay matching.

\medskip
\noindent\textbf{Step 2: (optional) phase separation.}
If a dynamical phase model is available (e.g.\ from characterised path-length differences and dispersion), compute $\Delta\phi^{\mathrm{dyn}}_j$ and subtract it from the measured $\Delta\theta_j$. The residual estimates $\Delta\gamma_j+\Delta\phi^{\mathrm{tech}}_j$. This step is not required for correcting the state, but it is important when one wishes to verify and use geometric phase as a control resource.

\medskip
\noindent\textbf{Step 3: diagonal feed-forward correction.}
Fixing a global reference $\vartheta_0=0$, build cumulative phases $\vartheta_{j+1}=\sum_{k=0}^{j}\Delta\theta_k$ and implement
\begin{equation}
  D_{\mathrm{corr}} = \mathrm{diag}\!\left(e^{-i\vartheta_0},\dots,e^{-i\vartheta_{d-1}}\right).
  \label{eq:feedforward}
\end{equation}
Applied to the measured state, $D_{\mathrm{corr}}$ cancels $U_{\mathrm{tot}}$ (up to a global phase), thereby restoring the target amplitudes.

\subsection{Measurement and tomography in $d$ dimensions}\label{subsec:tb-tomo}
The goal here is to specify which experimentally accessible fringes determine the density matrix elements and, simultaneously, to explain how intermediate diagnostics (e.g.\ fringe visibilities) quantify whether phase isolation is successful. In practice, a small set of pairwise scans is often sufficient for calibration, while additional scans provide cross-checks for consistency and coherence.

In addition to measuring the computational basis populations, it is useful to recover off-diagonal terms via interferometric mixing. For a general state with density matrix $\rho$, interference between bins $j$ and $k$ yields
\begin{equation}
P_{jk}(\varphi) = \tfrac{1}{2}(p_j+p_k)+\Re\!\left[e^{i\varphi}\rho_{jk}\right],
\label{eq:pair-fringe}
\end{equation}
where the fringe offset gives $\arg\rho_{jk}$ and the visibility gives $|\rho_{jk}|$ relative to $p_j,p_k$. A Fourier-basis measurement using an approximate $F_d$ implementation (Eq.~\eqref{eq:dft}) provides an independent consistency check: reconstructed $\rho$ should reproduce the measured Fourier-basis probabilities. Operationally, the same fitted parameters serve both as phase estimates (for feed-forward) and as coherence diagnostics (through visibility).

\subsection{Worked example: balanced two-stage tree ($m=2$) producing a qutrit}\label{subsec:qutrit-example}
This worked example shows, in the simplest nontrivial dimension, how adjacent-bin fringe offsets determine the correction matrix. The emphasis is on the measurement logic (recovering $\Delta\theta_j$ and checking coherence through visibility) rather than on algebraic manipulation.

For two cascaded balanced UMZIs ($\eta_1=\eta_2=\tfrac12$) producing $d=3$ bins, one obtains
\begin{align}
\alpha_1 &= \tfrac12, \nonumber \\
\alpha_2 &= \tfrac12(e^{i\phi_1}+e^{i\phi_2}), \\
\alpha_3 &= \tfrac12 e^{i(\phi_1+\phi_2)}. \nonumber
\label{eq:qutrit-amps}
\end{align}
Additional geometric and technical phases enter additively as $\phi_k\mapsto\phi_k+\delta\phi_k$ and therefore shift the observed fringe offsets. Adjacent-bin scans yield $\Delta\theta_1$ and $\Delta\theta_2$, from which one computes cumulative phases $\vartheta_1=\Delta\theta_1$ and $\vartheta_2=\Delta\theta_1+\Delta\theta_2$ (with $\vartheta_0=0$), and programs the per-bin phase shifter to apply $e^{-i\vartheta_j}$. A post-correction scan should show corrected offsets consistent with the intended target (e.g.\ near-zero offsets for a flat-phase reference) while maintaining high visibility, indicating successful phase isolation and compensation.

\section{Correction algorithms}\label{sec:correction}

The calibration of Sec.~\ref{sec:calibration} yields a consistent set of adjacent-bin relative phases $\Delta\theta_j=\theta_{j+1}-\theta_j$, which together encode the full phase budget (geometric, dynamical, and technical). The correction task is therefore to implement a diagonal feed-forward $D_{\mathrm{corr}}$ whose cumulative phases $\vartheta_{j+1}=\sum_{k=0}^{j}\Delta\theta_k$ cancel these offsets, restoring the intended target state up to a global phase. In experiments, $D_{\mathrm{corr}}$ is realised by programmable per-bin phase shifts (e.g.\ an EOM driven synchronously with the bin clock or a pulse shaper). The fitted visibilities from the calibration scans serve as intermediate diagnostics: if visibilities are low, the state is not phase-coherent across the corresponding bins and feed-forward cannot recover the intended pure-state phases.

For completeness, Algorithm~\ref{alg:berry-correction} sketches a numerical implementation in which the same phase budget is accumulated directly from a simulated state trajectory. The goal of the loop is to (i) accumulate the geometric contribution in a parallel-transport gauge via $\Im\langle\psi|\dot\psi\rangle$, (ii) accumulate the dynamical contribution via $-(1/\hbar)\langle\psi|H|\psi\rangle$, and (iii) apply a phase-only correction that subtracts the total from the evolving state.

\begin{figure}[t]
\captionof{algorithm}{Berry Phase Correction}\label{alg:berry-correction}
\begin{algorithmic}[1]
\Require state $\psi(t)$, $\mathrm{initial\_phase}$, field $B_{\mathrm{eff}}(t)$, step $\Delta t$, total time $T_{\mathrm{total}}$
\State $\mathrm{phase\_geom} \gets 0$
\State $\mathrm{phase\_dyn} \gets 0$
\State $\mathrm{phase\_total} \gets 0$
\While{circuit is running}
    \If{photon detected by sensor}
        \State $\dot{\psi}(t) \gets (\psi(t+\Delta t) - \psi(t))/\Delta t$
        \State $\mathrm{phase\_geom} \gets \mathrm{phase\_geom}$
        \Statex \hspace{\algorithmicindent}\hspace{\algorithmicindent}$+\,\Im\!\langle\psi(t)|\dot{\psi}(t)\rangle\,\Delta t$
        \State $H_t \gets \mathrm{Hamiltonian\_from\_field}(B_{\mathrm{eff}}(t))$
        \State $\mathrm{phase\_dyn} \gets \mathrm{phase\_dyn}$
        \Statex \hspace{\algorithmicindent}\hspace{\algorithmicindent}$-\,(1/\hbar)\,\Re\!\langle\psi(t)|H_t|\psi(t)\rangle\,\Delta t$
        \State $\mathrm{phase\_total} \gets \mathrm{phase\_geom} + \mathrm{phase\_dyn}$
        \State $\mathrm{state\_correct} \gets$
        \Statex \hspace{\algorithmicindent}\hspace{\algorithmicindent}$\exp[i(\mathrm{initial\_phase} - \mathrm{phase\_total})]\,\psi(t)$
        \State \textbf{output} $\mathrm{state\_correct}$
    \EndIf
    \State $\psi(t) \gets \mathrm{propagate\_state}(\psi(t), H_t, \Delta t)$
    \State $t \gets t + \Delta t$
\EndWhile
\end{algorithmic}
\end{figure}

\subsection{Correction algorithm outputs}

\begin{figure}[!ht]
    \centering
    {\bfseries Total phase $\beta(t)$.}\\[0.4em]
    \includegraphics[width=\linewidth]{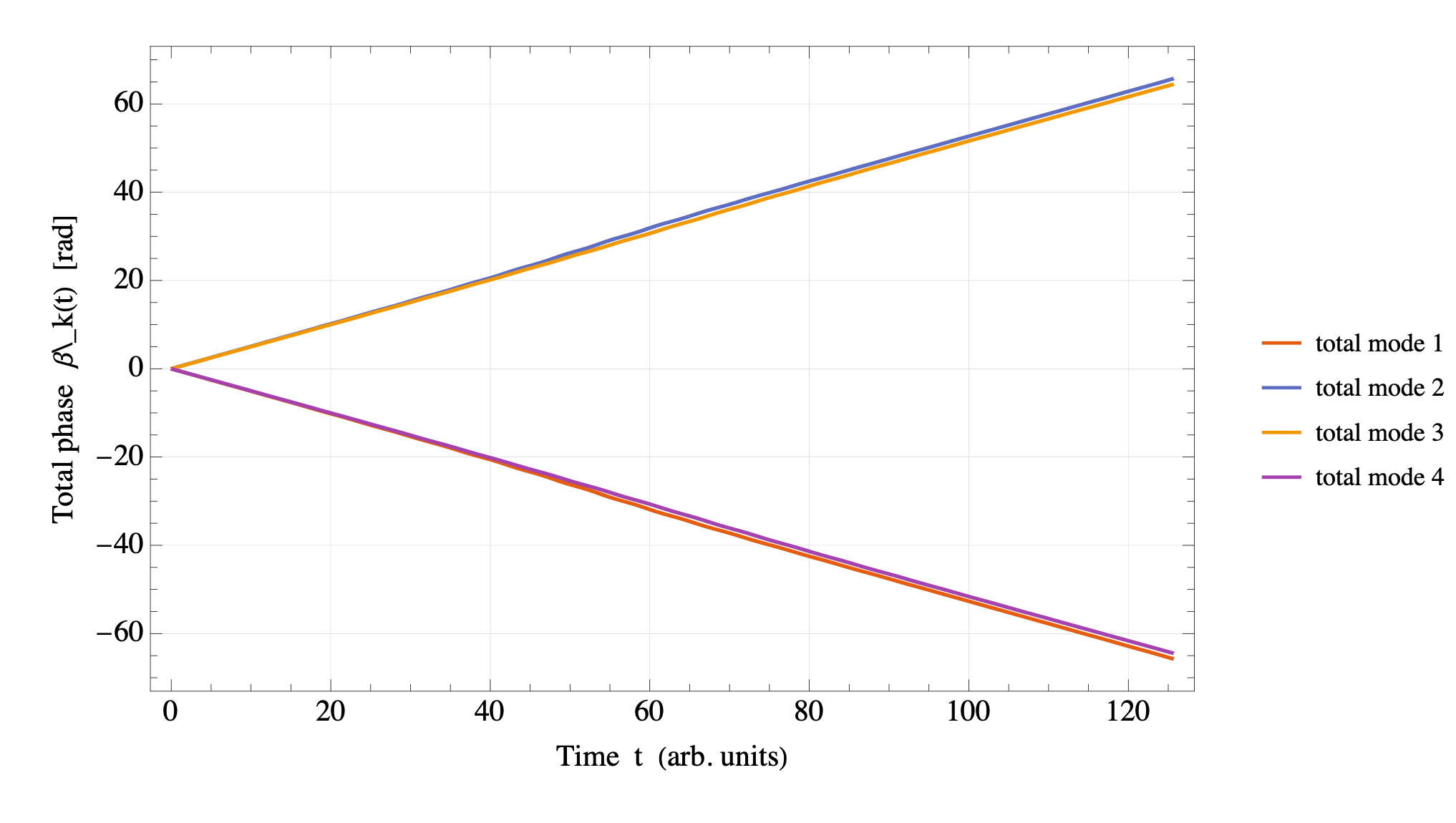}
    \caption{\textbf{Total phase} $\beta(t)$ (Pancharatnam phase) of the evolving state relative to the initial state. The decomposition $\beta=\gamma+\phi_{\rm dyn}$ holds within numerical precision.}
    \label{fig:total-phase}
\end{figure}

We study a 4-bin qudit formed by two decoupled spin-$\tfrac12$ systems, a convenient testbed for separating geometric and dynamical contributions. Figure~\ref{fig:total-phase} shows the total Pancharatnam phase $\beta(t)$ accumulated over the simulated evolution; the decomposition $\beta = \gamma + \phi_{\mathrm{dyn}}$ holds within numerical precision throughout.

\begin{figure}[!ht]
    \centering
    {\bfseries Dynamical phase $\phi_{\mathrm{dyn}}(t)$.}\\[0.4em]
    \includegraphics[width=\linewidth]{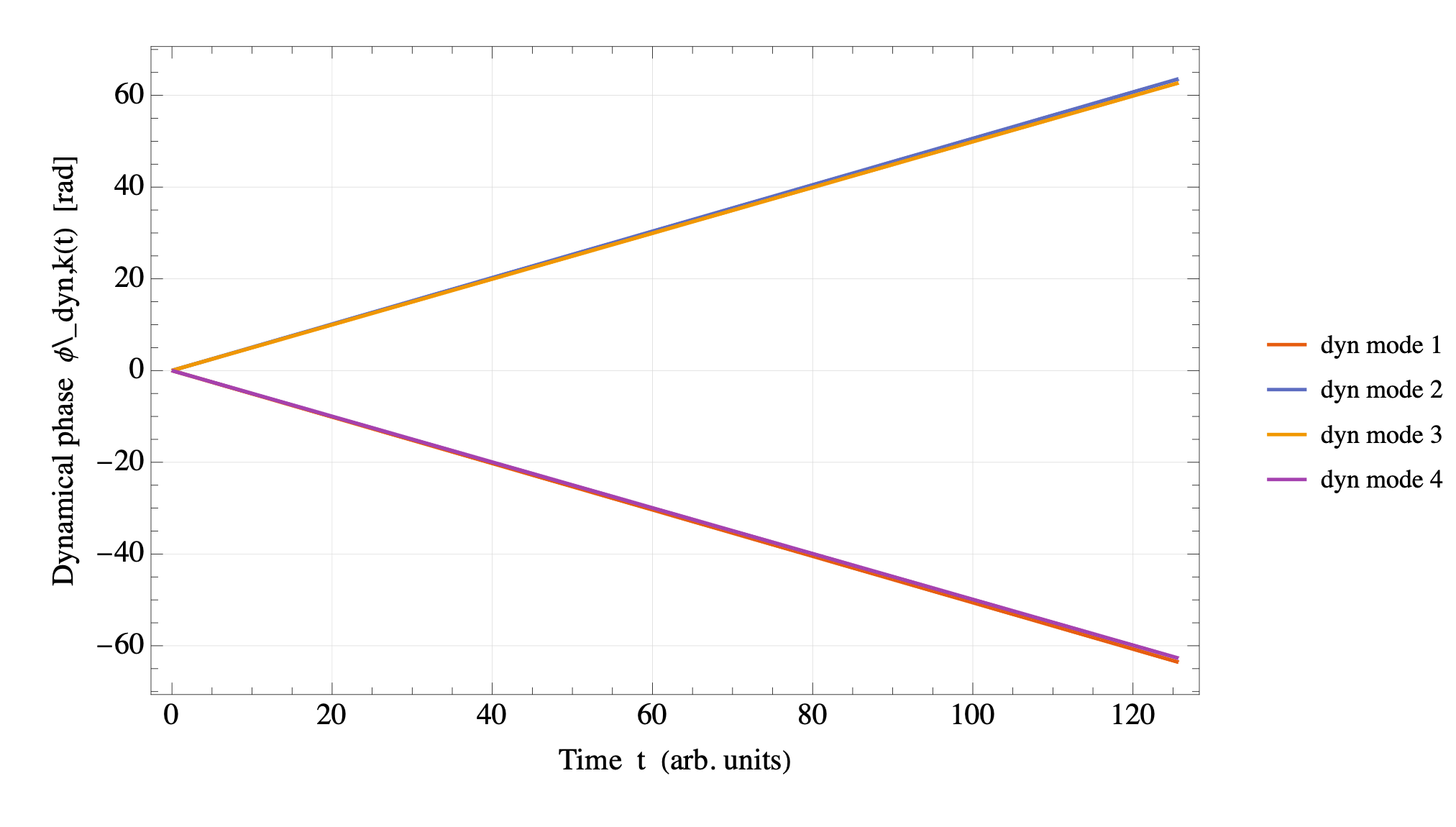}
    \caption{\textbf{Dynamical phase} $\phi_{\mathrm{dyn}}(t)$ extracted from the simulated evolution. This contribution reflects the time-integrated energy expectation and is path-dependent (not purely geometric).}
    \label{fig:dyn-phase}
\end{figure}

\begin{figure}[!ht]
    \centering
    {\bfseries Geometric phase $\gamma(t)$.}\\[0.4em]
    \includegraphics[width=\linewidth]{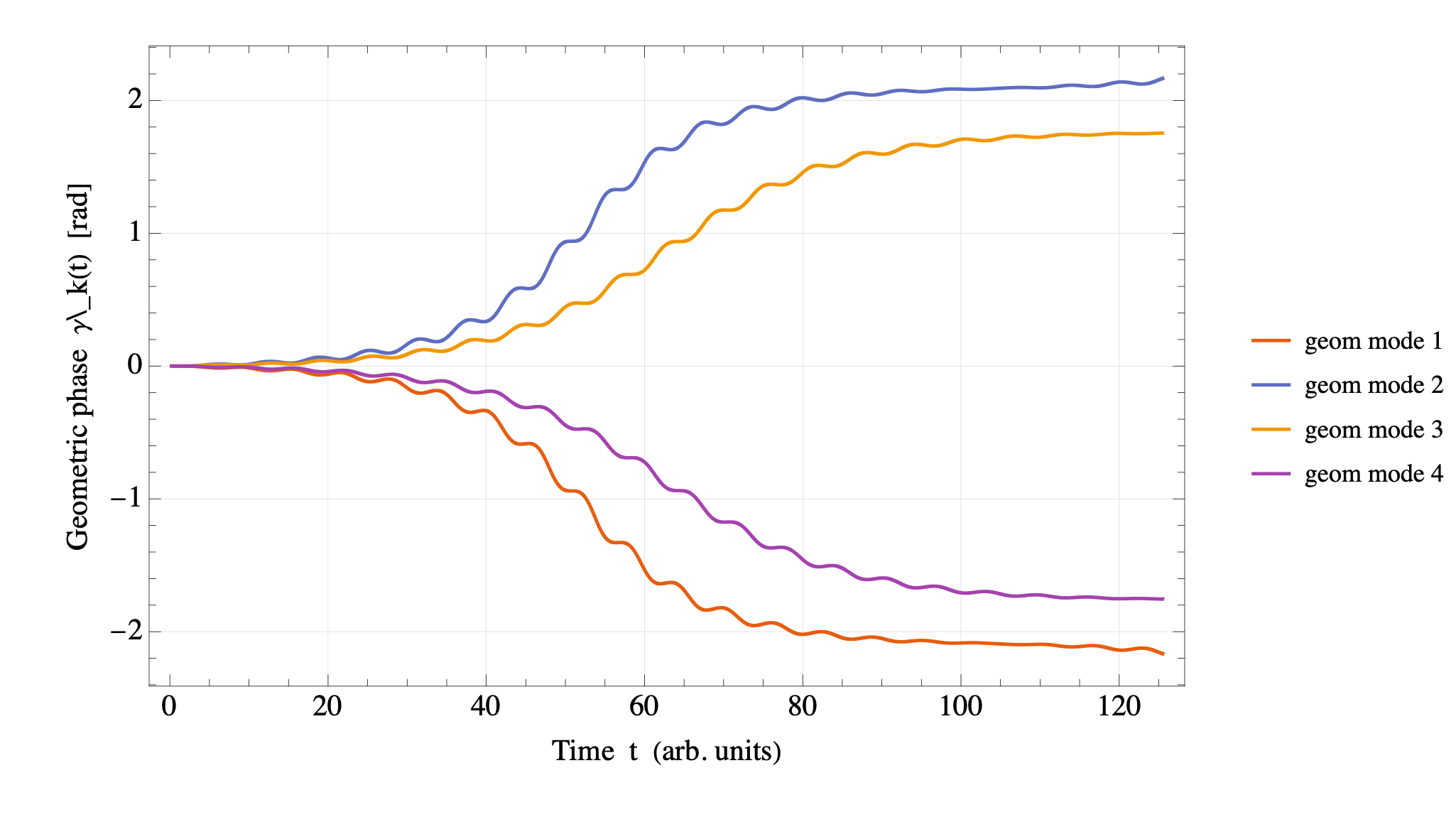}
    \caption{\textbf{Geometric phase} $\gamma(t)$ obtained after imposing a parallel-transport gauge and subtracting the dynamical contribution. This is the holonomy associated with the state trajectory.}
    \label{fig:geom-phase}
\end{figure}

The total phase separates cleanly into its dynamical and geometric contributions. Figure~\ref{fig:dyn-phase} shows the dynamical contribution $\phi_{\mathrm{dyn}}(t)$, which reflects the time-integrated energy expectation and grows monotonically with the evolution. Figure~\ref{fig:geom-phase} shows the geometric phase $\gamma(t)$ obtained after imposing the parallel-transport gauge and subtracting the dynamical contribution; this is the holonomy associated with the state trajectory and is path-dependent in the non-cyclic case.

\begin{figure}[!ht]
    \centering
    {\bfseries Per-bin total phases $\theta_j$ (absolute).}\\[0.4em]
    \includegraphics[width=\linewidth]{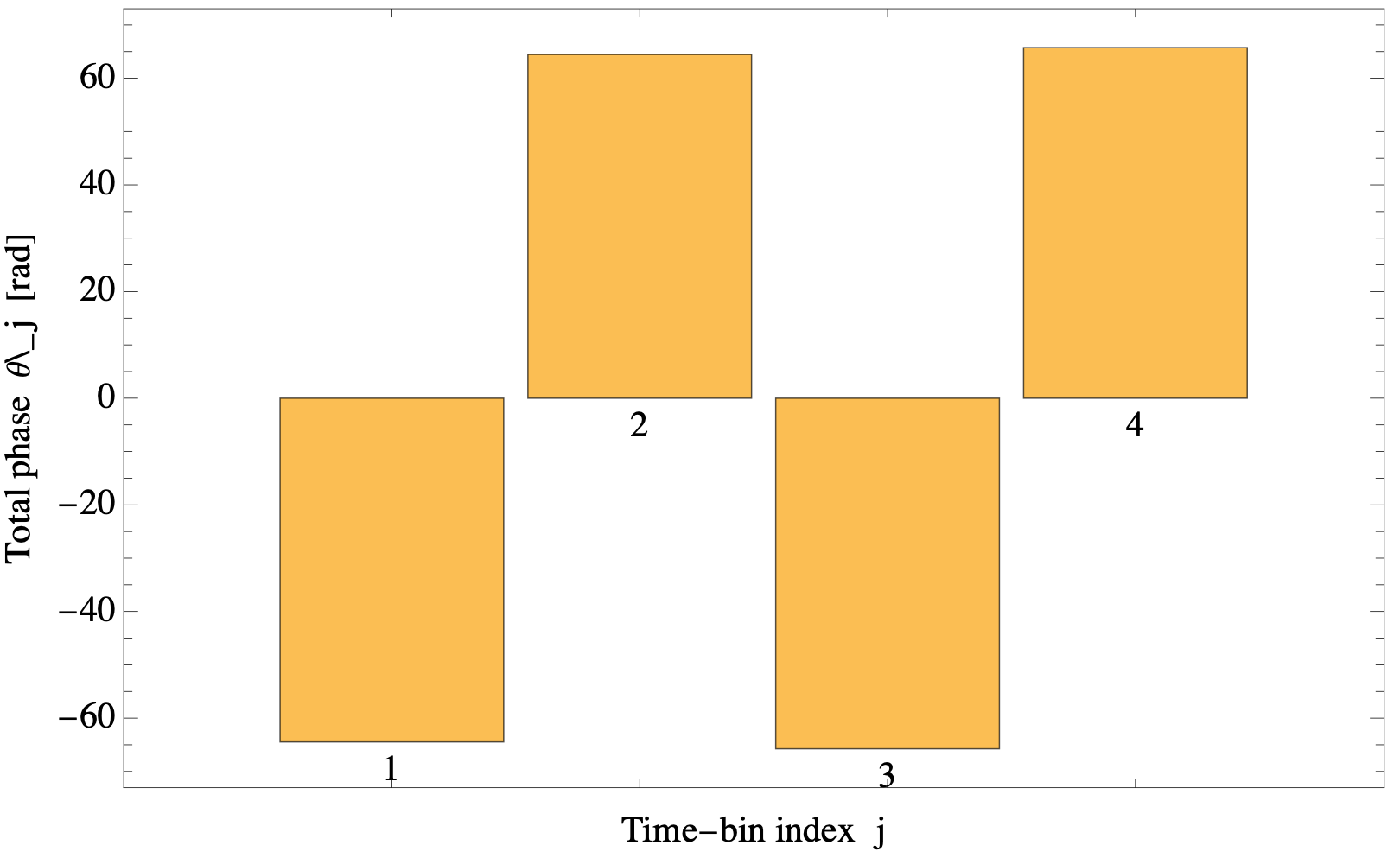}
    \caption{Extracted per-bin total phases $\theta_j$ before reducing modulo $2\pi$. The absolute values are gauge-dependent, but differences $\Delta\theta_j$ determine the physical relative phases between bins and define the correction matrix.}
    \label{fig:bin-abs}
\end{figure}

\begin{figure}[!ht]
    \centering
    {\bfseries Per-bin total phases $\theta_j$ (mod $2\pi$).}\\[0.4em]
    \includegraphics[width=\linewidth]{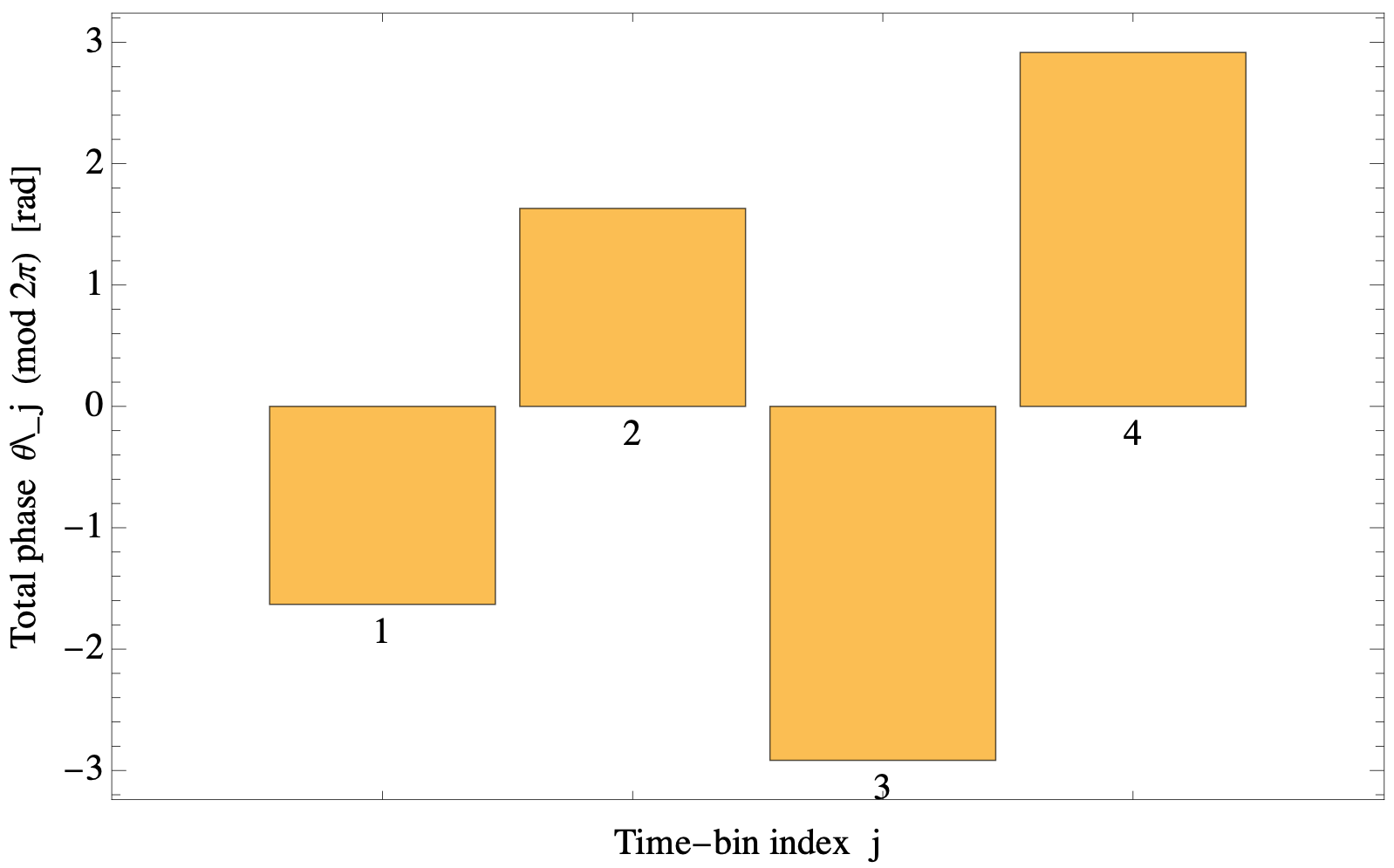}
    \caption{The modulo-$2\pi$ reductions reveal the interferometrically relevant relative phases; four distinct values reflect stable bin$\to$mode mapping and set the diagonal feed-forward $D_{\rm corr}=\mathrm{diag}(e^{-i\theta_j})$.}
    \label{fig:bin-mod}
\end{figure}

With the phase decomposition validated, we turn to the per-bin analysis that determines the feed-forward matrix $D_{\mathrm{corr}}$. Figure~\ref{fig:bin-abs} shows the extracted per-bin total phases $\theta_j$ in absolute form; the differences $\Delta\theta_j$ between bins are gauge-independent and define the correction. Figure~\ref{fig:bin-mod} shows these same phases reduced modulo $2\pi$, which are the interferometrically relevant quantities: four distinct values reflect a stable bin-to-mode mapping and set the diagonal $D_{\mathrm{corr}} = \mathrm{diag}(e^{-i\theta_j})$.

Taken together, the five figures validate the correction algorithm of Sec.~\ref{sec:correction}: the dynamical and geometric contributions are consistently separated, the reconstructed total phase matches their sum, and the inferred per-bin phases define a diagonal correction that restores the target bin phases up to a global phase. Minor deviations are consistent with finite numerical step size and the simplifying assumptions of lossless, dispersion-free propagation.

\paragraph{Practical laboratory workflow.}
In a laboratory setting, the protocol can be implemented with standard components already used for time-bin preparation and analysis. First, prepare the target $d$-bin state using the UMZI tree (or an equivalent source) and time-tag single-photon detections to obtain the computational-basis populations $p_j$. Second, insert an analysing UMZI with delay $\Delta t$ and, for each adjacent pair $(j,j{+}1)$, perform a routine sweep of the analyser phase $\varphi$ (using a piezo phase shifter or an EOM), record time-resolved single-photon counts in a fixed output port, and fit the sinusoid to extract the fringe offset (yielding $\Delta\theta_j$) and the visibility (a coherence check). Third, compute cumulative phases $\vartheta_{j+1}=\sum_{k=0}^{j}\Delta\theta_k$ with a chosen global reference, and program a per-bin phase shifter (EOM synchronized to the bin clock, pulse shaper, or equivalent) to apply $e^{-i\vartheta_j}$ on each bin. Finally, verify the correction by repeating one or two representative interference scans: successful correction is indicated by restored target fringe offsets (or near-zero residual offsets when the target is a flat-phase reference) while maintaining high visibility. Because all steps rely only on tunable UMZIs, routine phase sweeps, single-photon detectors, and programmable phase control, the workflow can be run periodically as a recalibration cycle without specialised hardware.

\section{Conclusion and Significance}

We establish a geometric-phase framework for time-bin photonic qudits and introduce a practical calibration protocol that isolates and compensates geometric (Pancharatnam--Berry), dynamical, and technical phases in interferometric time-bin circuits. By working directly in the time-bin basis and enforcing a parallel-transport gauge, we make geometric phase operational: it appears as a stable interferometric offset that can be extracted from routine phase sweeps and then cancelled through diagonal feed-forward phase control. A worked qutrit example with cascaded unbalanced interferometers illustrates how relative phases arise and how they are removed with our three-step procedure, while a multi-mode numerical case study demonstrates explicit separation of total, dynamical, and geometric components.

The implications are twofold. Theoretically, explicit geometric-phase extraction supports treating geometric phase as a controllable resource, enabling holonomic-style control strategies in temporal encodings. Practically, systematic phase separation and compensation strengthens phase stability for scalable quantum communication and photonic processing, where high-dimensional interference otherwise becomes increasingly sensitive to drift.

Next steps include benchtop validation with tunable UMZIs and electro-optic phase control, quantifying the impact of non-idealities (loss, residual dispersion, imperfect mode overlap) on both fringe visibility and phase-estimation accuracy, and studying scaling as $d$ increases beyond the near-term regime ($d\lesssim 10$). A further direction is integrating calibration outputs with error mitigation or error-correcting codes, where predictable phase structure can be leveraged to design more robust gates for quantum networks and distributed photonic processing.

\bibliographystyle{quantum}
\bibliography{references}

\end{document}